\documentclass[a4paper, dvipsnames, 12pt]{article}
\usepackage[warn]{mathtext}
\usepackage[utf8]{inputenc}
\usepackage{tikz}
\usepackage{graphicx}
\usepackage{amsmath,amsthm,amssymb}
\usepackage{alltt}
\usepackage{gensymb}
\usepackage[english]{babel}
\usepackage[top=1.5cm, left=2 cm]{geometry}
\usepackage{wrapfig}
\usepackage{float}
\usepackage{hyperref}
\usepackage{physics}
\usepackage{cancel}
\usepackage[english]{babel}
\usepackage{amsmath,amssymb,amsthm,mathtools,bm}
\usepackage{euscript,mathrsfs}
\usepackage{icomma}
\usepackage{graphicx}
\usepackage{wrapfig}

\usepackage{indentfirst}
\usepackage{amsbsy}
\usepackage{wasysym}
\usepackage{hyperref}
\usepackage{cancel}
\usepackage{verbatim}
\usepackage{empheq}
\usepackage{adjustbox}
\usepackage{authblk}
\usepackage{tikz-cd}

\usepackage{graphicx} 
\usepackage{amsfonts}
\usepackage{amsmath}
\usepackage{amssymb}
\usepackage{xcolor}
\usepackage{physics}
\usepackage{slashed}
\usepackage{tikz}
\usetikzlibrary{decorations.markings,calc,angles}
\tikzset{midarrow/.style={
    postaction = {decorate},
    decoration = {markings, mark = at position #1 with {\arrow{>}}}
    },
    midarrow/.default = 0.55
}

\newcommand{\ve}{\varepsilon}

\DeclareMathOperator{\arctanh}{artanh}

\begin{document}

\title{Composite worldline instantons and the nonperturbative particle decay in constant external
electric and magnetic fields}
\author{Alexander Gorsky$^{1}$ and Ivan Poluboyarinov$^{2}$}

\affil{$^1$Institute for Information Transmission Problems RAS, 127051 Moscow, Russia \\ 
$^2$ Moscow Institute of Physics and Technology, \\Institutsky lane 9, Dolgoprudny 141700, Russia\\

}

\maketitle

\begin{abstract}
    In this paper, we discuss the validity of the composite worldline instanton approach for the nonperturbative decay of charged particles in constant electric and magnetic fields. It is shown that the instanton results in the leading
    exponential approximation in the different limits  agree with the 
    ones obtained by the imaginary part of the self-energy in the external  field and by overlap of the wave functions. We comment on the possible application of our results for the estimation of the decay rate of the proton in the external 
    electric and magnetic field in the leading exponential approximation. The effects of entanglement of the particles in the final state are briefly mentioned.
    
\end{abstract}

\section{Introduction}
The Schwinger process of the pair creation in the external field 
is one of the simplest nonperturbative tunneling phenomenon in the QFT \cite{sauter1931verhalten,schwinger1951gauge}. There are a few ways to evaluate the probability of the process. 
The most familiar involves evaluation of the imaginary part 
of the effective one-loop action in the external field , see \cite{dunne2005heisenberg,fedotov2023advances} for the review.  A second related approach is based on the evaluation of the overlap of two exact wave functions of the emerging charged particles
in the constant electric field. 
In the third approach the probability is defined by the action evaluated on the  
worldline instantons in the Euclidean space-time \cite{affleck1982monopole,affleck1982pair}, see \cite{dunne2005worldline,dunne2006worldline,dumlu2010stokes,Dumlu_2011} for the more recent developments. 

The probability of pair creation can be increased by some  external stimulating factors, such as energetic photons. The evaluation of the probability in the time dependent electric field  is based on the Keldysh method. When the frequency of the external field is large enough  the unsuppressed multiphoton  creation of  pairs is the dominating
process \cite{brezin1970pair, popov1972pair}, see, for instance, \cite{monin2010photon, monin2010semiclassical,torgrimsson2017dynamically} for  recent  studies. The analogous processes of induced false vacuum decay have been discussed in \cite{affleck1979induced, voloshin1986particle, kuznetsov1997false, demidov2015high, gorsky2006particle}.

 The Schwinger processes have been discussed holographically in the
conformal gauge theory \cite{gorsky2002schwinger,semenoff2011holographic} or in the  holographic models of QCD \cite{sato2013holographic,hashimoto2015electromagnetic}.The critical electric field when the exponential suppression disappears has 
been found in the different holographic models. The entanglement of the created pairs has been evaluated and interpreted in terms of EPR bridges \cite{sonner2013holographic,jensen2013holographic}. The
 entanglement entropy of the pair was argued to coincide with the statistical Gibbs entropy of the particles created \cite{grieninger2023entanglement, grieninger2023entanglement2, ghodrati2015schwinger}.

A similar approach 
has been applied for the nonperturbative tunneling process of particle decay, which is energetically forbidden without the external field. Approaches based on the evaluation of the imaginary part of the self-energy in the external electric field or the overlap of the three exact wave functions in the external fields have been developed in \cite{ritus1988processes}.  
In this case, the worldline instantons are composites and involve several Euclidean trajectories of the daughter particles \cite{gorsky2002schwinger,monin2005monopole,gorsky2024}. 
The daughter particles can be of the same kind, say kink and antikink, or of the different kind, like dyon and monopole.
Such composite Euclidean trajectories have been previously discussed in 
false vacuum decay processes \cite{affleck1979induced,voloshin1986particle,gorsky2006particle} and in stimulated Schwinger processes \cite{monin2010photon,monin2010semiclassical}.  

Another example of complex worldline instantons has been considered in \cite{satunin2013width,satunin2015study} for the 
neutral massless particle decay in
 a constant magnetic field. The process of the photon decay in the magnetic field is known for a while \cite{robl1952pair,baier2007pair,klepikov1953emission,tsai1974photon} and in some domain of parameters it is exponentially suppressed. It was shown in  \cite{satunin2013width,satunin2015study} that the exponential factor for the decay of the neutral particle is reproduced by the 
complex composite worldline instanton,  the coordinate is complexified in this case.

In this paper, we focus on the nonperturbative decay of the charged particles in the external electric and magnetic field. We shall focus on the 
leading exponential suppression factors that do not take into account the spin effects. Hence a bit loosely we shall have in mind the
probability rate of process of proton decay in electric and magnetic fields.
The non-perturbative decay of the accelerated proton via the weak interaction 
has been suggested long ago in \cite{ginzburg1965pion}. However, the actual evaluation of the probabilities has been performed much later \cite{muller1997decay,PhysRevLett.87.151301,vanzella2000weak,matsas1999decay,wistisen2021transmutation} for constant  or periodic external field.
The energy required for the decay is gained from the electric field just as in the Schwinger process. 
The probability has been 
evaluated using the overlap of solutions to the Dirac equations
in the external field.

In this study we examine  the nonperturbative decay of a charged particle in a constant homogeneous electric field in the worldline formalism and compare it with  the evaluation of the imaginary part of the  self energy in the external electric field. The  coincidence of 
the results in the leading approximation confirms the validity of the worldline instanton approach. We have also checked the equivalence of the worldline instantons in the 
Euclidean and Rindler coordinate systems. In the magnetic field, we argue that the overlap of the wave functions fits the semiclassical result from the complexified worldline instanton. We briefly discuss the entanglement effects relevant for the correlation of transverse momenta of the daughter particles.

The paper is organized as follows. In Section 2 we describe the composite worldline instanton that describes the decay of the charged particle in the electric field to the final state of two particles. We argue that the evaluation of the action on the composite instanton in the Euclidean and Rindler coordinates yields the same answer.
In Section 3 we demonstrate that in the leading approximation the result derived by the worldline instanton coincides with the result derived from the imaginary part of the self-energy in the electric field. In Section 4 we briefly comment on the entanglement phenomena for the particles in the final state. In Section 5 we comment on the composite worldline instanton for the final state of three particles. The 
semiclassical approach based on the Routhian action is compared with the result derived from the wave function overlap
for the decay of the charged massive particle in the constant magnetic fields in Section 6. The results of the study are summarized in the Discussion, where the directions for  further research are outlined.

\section{Composite worldline instanton for charged particle decay}

In this Section we shall evaluate the decay probability rates in the
leading approximation using the generalization of the worldline
instanton approach suggested in \cite{affleck1982monopole,affleck1982pair}. We shall first focus on the decay 
of charged particles into charged and neutral ones in the constant external electric field. One could have in mind proton decay
mode $p\rightarrow n\pi^{+}$, although we consider only scalar particles for simplicity, since the spin factors do not influence the 
leading exponential factor. The composite worldline instanton configuration of worldlines of the particles can be determined in two natural 
frames; Euclidean and Rindler. In Rindler coordinates, the initial proton is at rest, while the neutron and charged meson 
evolve along the solution to their equation of motion. The whole instanton action involves two contributions, the weighted lengths of the worldlines of the particles  and the contribution of a closed Wilson loop whose shape is determined 
by the meson trajectory and the fictitious trajectory of the proton without decay. It can be represented as
\begin{equation} \label{eq:inst_action}
    S_\text{inst}=S_{\pi}+S_n -S_p
\end{equation}
where $S_i$ denotes the action of the corresponding particle trajectory. The $S_p$ is the action of the proton without decay
evaluated in the part of the trajectory between two interaction points.

\subsection{Rindler coordinates}
    We begin with the Rindler coordinate approach. Let us fix the gauge as $A_\mu = \begin{pmatrix} 0, & Et, & 0, & 0 \end{pmatrix}$ and perform transformation to Rindler coordinates:
    \begin{equation}
    	(t, x, y, z) \mapsto (\eta, \rho, y, z),
    \end{equation}
    \begin{equation}
    	t = \rho\sinh{\eta},\quad x = \rho\cosh\eta,
    \end{equation}
    and Wick rotation $\eta \mapsto -i\theta$. Varying the action of a charged particle, we obtain the following equation of motion:
    \begin{equation} \label{eq:EOM}
        \begin{gathered}
    	   \frac{d}{d\theta}\left(\frac{m\dot\rho}{\sqrt{\rho^2 + \dot\rho^2}} - qE\rho\sin\theta\cos\theta\right) - \\
           - \left(\frac{m\rho}{\sqrt{\rho^2 + \dot\rho^2}} + qE(2\rho\sin^2\theta - \dot\rho\sin\theta\cos\theta)\right) = 0.
        \end{gathered}
    \end{equation}
    
    In the Rindler coordinates, the initial proton is static. So, setting $\rho=R_p = \text{const}$ inside \eqref{eq:EOM}, we obtain
    \begin{equation} \label{eq:initial_particle_minus}
    	R_p = \frac{m_p}{|qE|}.
    \end{equation}
    For the trajectory of the neutral particle, we have to set $q=0$, and the corrsponding trajectory is
    \begin{equation}
    	\rho(\theta) = \frac{d}{\cos(\theta - \alpha)}
    \end{equation}
    for arbitrarily $d$ and $\alpha$.

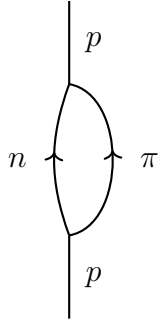
\begin{figure}[h!]
        \centering
        \begin{tikzpicture}[thick,scale=1]
            \coordinate (T) at (0,1);
            \coordinate (B) at (0,-1);

            \draw (T) -- ++(0,1.1) node[midway, right=2pt] {$p$};
            \draw (B) -- ++ (0,-1.1) node[midway, right=2pt] {$p$};

            \draw[midarrow] (B) to[out=110,in=-110] node[midway,left=6pt] {$n$} (T);
            \draw [midarrow] (B) to[out=10,in=-10] node[midway,right=6pt] {$\pi$} (T);
            
        \end{tikzpicture}
        \caption{Bounce for proton decay in Rindler coordinates}
        \label{fig:bounce_rindler}
    \end{figure}

    For a daughter charged particle trajectory, we can find that \eqref{eq:EOM} admits the solution
    \begin{equation} \label{eq:pion_traj}
    	\rho(\theta) = U(\theta) \pm \sqrt{U(\theta)^2 - \Gamma}, \quad
    	U(\theta) = A\cos\theta + B\sin\theta,\quad
    	\Gamma = A^2 + B^2 - r_\pi^2
    \end{equation}
    with $r_\pi=\frac{m_\pi}{|qE|}$ for arbitrary $A$ and $B$. One can note that \eqref{eq:pion_traj} is simply a circle with radius $r_\pi$ in Euclidean coordinates written in  Rindler coordinates. The curve can be written in parametric form as
    \begin{equation}
        \rho(\phi)\cos\theta(\phi) = A + r_\pi\cos\phi, \qquad \rho(\phi)\sin\theta(\phi) = B + r_\pi\sin\phi. 
    \end{equation}
    The composite bounce is shown in Fig. \ref{fig:bounce_rindler}.

    The parameters $d,\alpha,A,B$ are fixed by requiring that all three segments meet at two junction points and that momentum be conserved at each junction, which yields the corresponding equations:

    \begin{equation}
        \begin{cases}
            R_p = \frac{d}{\cos(\theta_0 - \alpha)},\\
            R_p \cos\theta_0 = A + r_\pi\cos\phi_0,\\
            R_p\sin\theta_0 = B + r_\pi\sin\phi_0,\\
            m_n\sin(\theta_0 - \alpha) + m_\pi\sin(\theta_0 - \phi_0) = 0,\\
            -m_p + m_n\cos(\theta_0 - \alpha) + m_\pi\cos(\theta_0-\phi_0) = 0.
        \end{cases}
    \end{equation}

    One can numerically solve this system of equations and evaluate \eqref{eq:inst_action}. As we shall see in the next section, the results agree with the analytical  expression in Euclidean coordinates\eqref{eq:instanton_action}. As analytical derivation in Euclidean coordinates is more compact and transparent, we present it in the next section, and use the Rindler configuration only as a numerical cross-check.

    \subsection{Bounce in Euclidean coordinates}\label{sec:cartesian_coordinates}
    We now turn to Euclidean coordinates and derive an analytical formula for the instanton action. In these coordinates, the initial proton is accelerated, so the Wick rotation is less direct than in the Rindler description. Hence, we will use Euclidean construction primarily as a shortcut to the analytical result, and the agreement with Rindler-coordinate approach will be a consistency check.

    As earlier, we start with the classical action for charged particles and perform Wick rotation:
    \begin{equation}
    	S = -m\int ds + q\int A_\mu dx^\mu,
    	\qquad
    	A_\mu =
    	\begin{pmatrix}
    		-Ex, & 0, & 0, & 0
    	\end{pmatrix}^T.
    \end{equation}
    \begin{equation} \label{eq:cartesian_action}
    	S_E = -iS =	m\int dt_E\sqrt{1+\dot x^2 + \dot y^2 + \dot z^2}	+ qE\int x\,dt_E.
    \end{equation}
    
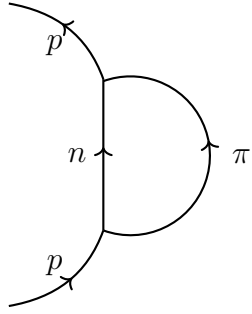
\begin{figure}[h!]
    	\centering
    	\begin{tikzpicture}[thick,scale=1]
            \coordinate (T) at (0,1);
            \coordinate (B) at (0,-1);
            \coordinate (LT) at (-1.25,2);
            \coordinate (LB) at (-1.25,-2);

            \draw[midarrow] (B) to node[midway,left=2pt]{$n$} (T);
            \draw[midarrow] (LB) to [out=10,in=-110] node[pos=0.7, left=4pt] {$p$} (B);
            \draw[midarrow] (T) to [out=110,in=-10] node[pos=0.3,left=4pt] {$p$} (LT);

            \draw[midarrow] (B) arc[start angle=-110, end angle=110, radius=1.05] node[midway,right=4pt] {$\pi$};
            
        \end{tikzpicture}
    	\caption{Bounce for proton decay in Euclidean coordinates}
    	\label{fig:bounce_cartesian}
    \end{figure}

    Assuming $\dot y = \dot z = 0$, we find the trajectories as functions of proper time
    \begin{equation}
    	t_E(\tau) = R\sin(\omega\tau+\alpha_0),\qquad
    	x(\tau) = R\cos(\omega\tau+\alpha_0), \qquad
    	R = \frac{m}{|qE|}.
    \end{equation}
    rescaling the proper time, we set $\omega = 1$, $\alpha_0 = 0$. The result is well known: charged particles are in cyclotron motion. The bounce has the shape shown in Fig. \ref{fig:bounce_cartesian}.

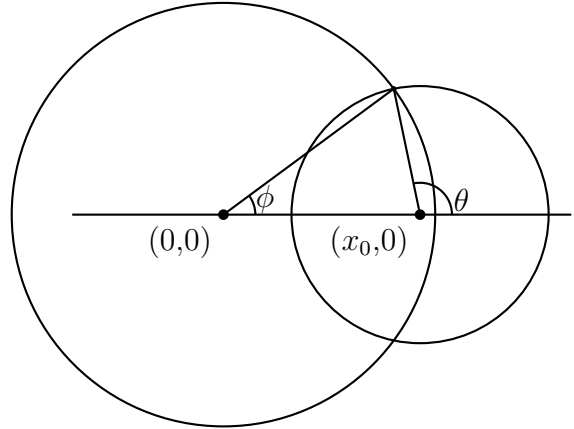
\begin{figure}[h!]
        \centering
        \begin{tikzpicture}[thick,scale=1]
            \def\R{2.8}
            \def\r{1.7}
            \def\x{2.6}

            \pgfmathsetmacro{\a}{(\R*\R - \r*\r + \x*\x)/(2*\x)}
            \pgfmathsetmacro{\h}{sqrt(\R*\R - \a*\a)}

            \coordinate (O1) at (0,0);
            \coordinate (O2) at (\x,0);
            \coordinate (P) at (\a,\h);
            \coordinate (X1) at ($(O1) - (2,0)$);
            \coordinate (X2) at ($(O2) + (2,0)$);

            \draw (O1) circle (\R);
            \draw (O2) circle (\r);
            \draw (O1) -- (P);
            \draw (O2) -- (P);
            \draw (X1) to (X2);

            \fill (O1) circle (2pt) node[below left] {$(0,0)$};
            \fill (O2) circle (2pt) node[below left] {$(x_0,0)$};
            \fill (P) circle (1pt);

            \pic[draw, angle radius=12pt, angle eccentricity=1.25]{angle = X2--O1--P};
            \pic[draw, angle radius=12pt, angle eccentricity=1.25]{angle = X2--O2--P};
            \node at ($(O2)+(0.55,0.20)$) {$\theta$};
            \node at ($(O1)+(0.55,0.20)$) {$\phi$};
            
        \end{tikzpicture}
        \caption{Geometry used for junction equations} \label{fig:cartesian_geometry}
    \end{figure}

    Once again, we have to satisfy two conditions: trajectories intersect in two points, and the momentum conservation law is satisfied. Set up the following notation. Let two circles (the trajectories of the initial particle and the charged daughter particle) intersect at points corresponding to proper times $\pm\phi$ for the initial particle and $\pm\theta$ for the daughter particle. The center of the daughter circle is shifted along the $x$ axis by $x_0$ (see Fig. \ref{fig:cartesian_geometry}).
    
    Then conditions become:
    \begin{equation}
    	\begin{cases}
    		R_p\sin\phi = R_\pi\sin\theta,\\
    		R_p\cos\phi = x_0 + R_\pi\cos\theta,\\
            m_p\cos\phi = m_\pi\cos\theta + m_n.  
    	\end{cases}
    \end{equation}
    Solving this system, we find the following result
    \begin{equation}
    	\theta = \arccos(\Gamma),\qquad
    	\Gamma = \frac{m_p^2-m_\pi^2-m_n^2}{2m_\pi m_n},
    \end{equation}
    \begin{equation}
    	\phi = \arcsin \left(\frac{m_\pi}{m_p}\sqrt{1-\Gamma^2}\right),
    \end{equation}
    \begin{equation}
    	x_0 = -\frac{1}{qE}	\left[m_p\sqrt{1-\left(\frac{m_\pi}{m_p}\sqrt{1-\Gamma^2}\right)^2}	- m_\pi\Gamma\right].
    \end{equation}
    Substituting these trajectories into actions \eqref{eq:cartesian_action} and combining them into \eqref{eq:inst_action}, we get
    \begin{equation} \label{eq:instanton_action}
        \begin{gathered}
            S_\text{inst} = \frac{m_\pi^2}{|eE|}\arccos\left(\frac{m_p^2 - m_\pi^2 - m_n^2}{2m_\pi m_n}\right) - \frac{m_p^2}{|eE|}\arccos\left(\frac{m_p^2 + m_n^2 - m_\pi^2}{2m_p m_n}\right) +\\
            +\frac{m_\pi m_n}{|eE|}\sqrt{1-\left(\frac{m_\pi^2 + m_n^2 - m_p^2}{2m_\pi m_n}\right)^2}.
        \end{gathered}
    \end{equation}
    
    As was stated earlier, we can justify this formula by numeric comparison with results obtained by the Rindler coordinates approach. Indeed, direct calculation shows that agreement is exact within numerical precision.

    Now we should discuss the agreement of this formula with the results obtained in \cite{ritus1988processes} earlier. There the results were obtained in the following two approximation regimes: the case of equal masses of  charged particles  and slightly different masses of charged particles. We will show that \eqref{eq:instanton_action} coincides with the results in \cite{ritus1988processes}.

    Firstly, consider the case of equal masses of charged particles. Setting $m_\pi = m_p = m, m_n = \mu$, we obtain the following.
    \begin{equation}
        \frac{m_p^2 - m_\pi^2 - m_n^2}{2m_\pi m_n} = -\frac{\mu}{2m} =: -\xi,\qquad \frac{m_p^2 + m_n^2 - m_\pi^2}{2m_p m_n} = \frac{\mu}{2m} = \xi.
    \end{equation}
    therefore,
    \begin{equation}
        S_\text{inst} = \frac{1}{eE}\Big[m^2 \arccos(-\xi) - m^2\arccos(\xi) + m\mu\sqrt{1-\xi^2}\Big].
    \end{equation}
    Taking into account that $\arccos(-\xi) = \pi - \arccos(\xi)$ and $\arcsin(x)+\arccos(x) = \frac{\pi}{2}$ we arrive at 
    \begin{equation}
        S_\text{inst} = \frac{2m^2}{eE}\Big[\arcsin\xi + \xi\sqrt{1-\xi^2}\Big],\qquad \xi = \frac{\mu}{2m},
    \end{equation}
    which exactly reproduces the result of \cite{ritus1988processes}.

    Now, consider the second regime, where the masses of charged particles differ  slightly: $0<\Delta :=m_\pi-m_p \ll m_\pi,m_p$. Define $ m=m_p,m_\pi=m+\Delta,m_n=\mu$ and also $\mu/m \ll 1$. Given this, we have
    \begin{equation}
        \begin{gathered}
        \frac{m_p^2 - m_\pi^2 - m_n^2}{2m_\pi m_n} = -\frac{\Delta}{\mu}-\frac{\mu^2 - \Delta^2}{2m\mu} + \mathcal O(m^{-2}),\\
        \frac{m_p^2 + m_n^2 - m_\pi^2}{2m_p m_n} = -\frac{\Delta}{\mu}+\frac{\mu^2 - \Delta^2}{2m\mu}.
        \end{gathered}
    \end{equation}
    Defining $\omega_0 = \frac{\sqrt{\mu^2 - \Delta^2}}{\Delta}$, define
    \begin{equation}
        x = -\frac{\Delta}{\mu} = -\frac{1}{\sqrt{1+\omega_0^2}},\qquad \ve = \frac{\mu^2 - \Delta^2}{2m\mu} = \frac{\Delta}{2m}\frac{\omega_0^2}{\sqrt{1+\omega_0^2}} \ll 1,
    \end{equation}
    where the last inequality holds due to $\mu \ll m$. Given this, instanton action can be written as
    \begin{equation}
        \begin{gathered}
            S_\text{inst} = \frac{1}{eE}\left[2m\Delta\arccos(x) + m^2\frac{2\ve}{\sqrt{1-x^2}} + m\Delta\omega_0 + \mathcal O(\Delta^2)\right].
        \end{gathered}
    \end{equation}
    Expressing everything through $\omega_0$:
    \begin{equation}
        \frac{2\ve}{\sqrt{1-x^2}} = \frac{\Delta}{m}\omega_0,\qquad \arccos\left(-\frac{\Delta}{\mu}\right) = \pi - \arctan\omega_0,
    \end{equation}
    for action, we finally have:
    \begin{equation}
        S_\text{inst} = \frac{2m\Delta}{eE}\Big[\pi + \omega_0 - \arctan\omega_0\Big],
    \end{equation}
    which completely matches the results obtained in \cite{ritus1988processes}.

    Now we shall discuss the physically relevant limit of the light charged daughter particle. Denote $\delta = m_n - m_p$, replace $m_\pi$ by $m_c$ (which stands for "charged") for more generality and consider the limit $m_c \ll m_n,m_p$. Then, the action can be approximately written as
    \begin{equation}
        S_\text{inst}\approx \frac{1}{|eE|}\left[ m_c^2 \arccos\left(-\frac{\delta}{m_c}\right) + \delta\sqrt{m_c^2 - \delta^2}\right].
    \end{equation}
    An interesting effect can be seen from this formula. For the case of a very light charged daughter particle $m_c < \delta$, this formula is no longer valid in the real domain and requires additional clarification, which will be provided in the following. 

    Let us first start our discussion with the real $m_c \geq \delta$ regime. There two additional interesting subregimes arise: $m_c \gg \delta$ (which corresponds to the actual $p\rightarrow n\pi$ decay), and "real threshold" limit $m_c\rightarrow \delta+0$. Expanding the instanton action according to the first one, we obtain the following.
    \begin{equation}
        S_\text{inst} = \frac{1}{|eE|}\left( \frac{\pi m_c^2}{2} + 2m_c (m_n-m_p) \right)  \approx \frac{\pi m_c^2}{2|eE|}.
    \end{equation}
    Note that the factor of 2 in the denominator has a transparent physical meaning. Indeed, in order to satisfy the momentum conservation law at the junction points given the heavy charged daughter, its trajectory should be almost perpendicular to the neutron trajectory. So, the charged daughter trajectory is almost half that of a circle. Recall that the instanton action for the Schwinger effect is $S_\text{inst}^\text{Schwinger} = \frac{\pi m_c^2}{|eE|}$, and the combined trajectory of particles forms a circle. In our case, we have only half a circle, and the additional 2 in the denominator is a manifestation of this fact. 
    
    It is also interesting to compare this result with the neutron decay result obtained in \cite{gorsky2024}. Expanding the instanton action for the neutron decay from \cite{gorsky2024} within the same limits, we obtain
    \begin{equation}
        S_\text{inst}^\text{neutron} = \frac{1}{|eE|}\left( \frac{\pi m_c^2}{2} - 2m_c (m_n-m_p) \right).
    \end{equation}
    Note the opposite sign of the term linear in $m_c$. This sign is responsible for the fact that the decay of a proton is suppressed more than the decay of a neutron by an amount proportional to the difference in the masses of these particles, which is in full compliance with  intuitive expectations.

    In the $m_c \rightarrow \delta+0$ limit, we have
    \begin{equation}
        S_\text{inst} \rightarrow \frac{\pi m_c^2}{|eE|}.
    \end{equation}
    Again, this answer has a transparent physical meaning. The angle at which the charged daughter particle flies out of the junction is determined by the momentum conservation law. When the mass of this particle becomes almost equal to the difference between the masses of the neutron and proton, the only way to carry away a sufficient amount of momentum is to start moving in the direction almost opposite to the direction of motion of the proton. Thus, the trajectory of the charged daughter particle is almost an entire circle, and we arrive at a value that coincides with the value of the action in the Schwinger effect.

    Now we are ready to consider the case $m_c < \delta$. The reason why the instanton action is no longer real can be understood directly from the geometry of the bounce. As was explained in the previous paragraphs, given that the mass of the neutron is greater than the mass of the proton, the charged daughter particle has to carry away enough momentum from the junction point so that the momentum conservation law can be satisfied. For $m_c < \delta$, this is not possible for real trajectories. The real instanton therefore ceases to exist, the saddle point of the Euclidean action becomes complex, and the tunneling suppression is determined by the real part of the action. We consider the analytical continuation of formula \eqref{eq:instanton_action}, and then take the real part of the action. Denote
    \begin{equation}
        \theta = \arccos\frac{m_p^2 - m_c^2 - m_n^2}{2m_cm_n},\qquad \phi = \arccos\frac{m_p^2 + m_n^2 - m_c^2}{2m_p m_n},
    \end{equation}
    taking into account that
    \begin{equation}\label{eq:values_of_cosines}
        \frac{m_p^2 - m_c^2 - m_n^2}{2m_cm_n} < -1,\qquad \frac{m_p^2 + m_n^2 - m_c^2}{2m_p m_n} > 1,
    \end{equation}
    both angles are complex: $\theta = \theta_1 + i\theta_2, \phi = \phi_1 + i\phi_2$. So we have
    \begin{equation}
        \cos\theta = \cos\theta_1\cosh\theta_2 - i\sin\theta_1\sinh\theta_2 = \frac{m_p^2 - m_c^2 - m_n^2}{2m_cm_n}\in \mathbb R,
    \end{equation}
    The right hand side is real, hence $\sin\theta_1 = 0,\sin\phi_1 = 0$. Taking into account the necessary signs of the expressions from  \eqref{eq:values_of_cosines}, we get
    \begin{equation}
        \theta = (2k+1)\pi \pm i\eta,\qquad \phi = 2l\pi \pm i\xi,\qquad \eta,\xi>0. 
    \end{equation}
    The junction equation $m_p\sin\phi = m_c\sin\theta$ additionally fixes the signs in front of the imaginary parts of the angles. 
    \begin{equation}
        \theta = (2k+1)\pi \mp i\eta,\qquad \phi = 2l\pi \pm i\xi,\qquad \eta,\xi>0.
    \end{equation}
    Analytical continuation of the minimal bounce, without extra windings, fixes $k=l=0$, and substituting these angles inside \eqref{eq:instanton_action}, we obtain
    \begin{equation}
        S_\text{inst} = \frac{\pi m_c^2}{|eE|} + i\Phi(m_p,m_n,m_c),\qquad \Phi\in\mathbb R.
    \end{equation}
    This is the natural minimal analytic continuation of \eqref{eq:instanton_action}, continuously connected to the real bounce at $m_c = \delta$. Also note that we have not made any approximations here, so, in the case of $m_c < m_n - m_p$, the real part of the instanton action and hence the tunneling exponential suppression is determined solely by the mass of the charged daughter particle and does not depend on the masses of initial and neutral particles.

    \section{Decay rate  from the imaginary part of the self-energy}
    In this Section we use the imaginary part of the proton self-energy in the external field as an independent check of the instanton results. It is known that the decay rate can be calculated from the imaginary part of the self energy:
    \begin{equation} \label{eq:decay_rate_via_imaginary_part}
        \Gamma_p \propto \bra{p,\text{in}}\Im\Sigma_E\ket{p,\text{in}}.
    \end{equation}
    We do not attempt here a complete computation of \eqref{eq:decay_rate_via_imaginary_part}, which would require projecting $\Im\Sigma$ onto the nontrivial accelerated proton state. Instead, our goal is to extract the leading exponential suppression from $\Im\Sigma$ in the weak-field regime. We show that this asymptotic behavior reproduces the instanton exponent derived earlier.
    
    For simplicity, we will treat all particles as scalar ones, as it is sufficient for the leading exponential accuracy.  The self-energy is given (in Euclidean space) by
    \begin{equation}
        \Sigma(p) = g^2\int\frac{d^4 k}{(2\pi)^4}G_n(k)G_\pi(p-k;E),
    \end{equation}
    and Euclidean propagators in external field can be written in Schwinger proper time form:
    \begin{equation}
        G_n^E(k) = \frac{1}{k_\parallel^2 + k_\perp^2 + m_n^2} = \int_0^\infty ds_n e^{-s_n(m_n^2 + k_\parallel^2 + k_\perp^2)},
    \end{equation}
    \begin{equation}
        G_\pi^E(p;E) = \int_0^\infty ds_\pi \frac{1}{\cos(|eE|s_\pi)}e^{-s_\pi(m_\pi^2 + p_\perp^2) - \frac{\tan(|eE|s_\pi)}{|eE|}p_\parallel^2},
    \end{equation}
    where $p_\parallel = (p_\tau,p_x),p_\perp=(p_y,p_z)$. Setting $p_\perp = 0$, self energy is given by
    \begin{equation}
        \begin{gathered}
            \Sigma = g^2 \int ds_n\int ds_\pi\int \frac{d^4k}{(2\pi)^4}\frac{1}{\cos(|eE|s_\pi)} \times \\
            \times\ e^{-s_n(m_n^2 + k^2) -s_\pi(m_\pi^2 + k_\perp^2) -\frac{\tan(|eE|s_\pi)}{|eE|}(p_\parallel - k_\parallel)^2}.
        \end{gathered}
    \end{equation}
    The propagator contour prescription $\pm i0$ shifts the $s_\pi$ slightly above/below the real axis. Consider the following calculation, which shows that the imaginary part of $\Sigma$ is determined by the residues at the Schwinger poles:
	\begin{equation}
        \begin{gathered}
		      \Im \Sigma = \frac{1}{2i}(\Sigma_{s-i0} - \Sigma_{s-i0}^\dagger) = \frac{1}{2i}(\Sigma_{s-i0} - \Sigma_{s+i0}) = \\
              = \frac{1}{2i}2\pi i \sum \Res \Sigma = \pi\sum\Res\Sigma.
        \end{gathered}
	\end{equation}
	The singularities are given by the solutions of $\cos(|eE|s_\pi) = 0$, so we have to calculate the residues at $|eE|s_\pi = \frac{\pi}{2}(2l+1),\ l=0,1,\dots$. Note that due to the $\tan(|eE|s_\pi)$ in the exponent, this is an essential singularity and must be handled through the Laurent expansion. In particular, substitute $|eE|s_\pi = \frac{\pi}{2}(2l+1) + z$ and expand trigonometric functions around $z\ll1$:
    \begin{equation}
        \cos(|eE|s_\pi) = (-1)^{l+1}\sin z,\qquad \tan(|eE|s_\pi) = -\cot(z).
    \end{equation}
    The imaginary part of $\Sigma$ is then given by
    \begin{equation}
        \begin{gathered}
            \Im\Sigma = \sum_{l=0}^\infty (-1)^{l+1}\frac{\pi g^2}{|eE|}\int ds_n \int\frac{d^4 k}{(2\pi)^4}e^{-s_n(m_n^2 + k^2)}e^{-\frac{\pi}{2eE}(2l+1)(m_\pi^2 + k_\perp^2)} \times \\
            \times \Res_{z=0}\Big[\frac{1}{\sin z} e^{\frac{(p_\parallel - k_\parallel)^2}{|eE|}\cot z - \frac{m_\pi^2 + k_\perp^2}{|eE|}z}\Big].
        \end{gathered}
    \end{equation}
    Introducing notation $a=|eE|,q=p_\parallel - k_\parallel$, summing the geometric series, and introducing notation as in \cite{ritus1988processes}:
    \begin{equation}
        \nu = \frac{m_p^2}{2|eE|},\quad \nu' = \frac{m_\pi^2 + k_\perp^2}{2|eE|},\quad \rho = \frac{m_n^2 + k_\perp^2}{2|eE|},
    \end{equation}
    we have
    \begin{equation} \label{eq:imaginary_sigma_intermediate}
        \begin{gathered}
            \Im\Sigma = -\int d^2k_\perp \frac{\pi g^2}{a}\frac{e^{-\pi\nu'}}{1+e^{-2\pi\nu'}}\int ds_n \times \\
            \times \int \frac{d^2q}{(2\pi)^4}e^{-s_n(2a\rho + (p-q)^2)}\Res_{z=0}\Big[\frac{1}{\sin z}e^{\frac{q^2}{a}\cot z - 2\nu'z}\Big].
        \end{gathered}
    \end{equation}
    The residue is given by
    \begin{equation}
        \Res_{z=0}\Big[\frac{1}{\sin z}e^{A\cot z - Bz}\Big] = e^{-iA}{}_1F_{1}(\frac{1+iB}{2},1,2iA),
    \end{equation}
    Equation \eqref{eq:imaginary_sigma_intermediate} can be simplified to
    \begin{equation} \label{eq:imaginary_part_final_integral}
        \Im\Sigma = -\int dk \frac{kg^2}{8\pi a}\frac{e^{-2\pi\nu'}}{1+e^{-2\pi\nu'}}\int_0^\infty d\tau\ e^{-2(\rho+\nu)\tau + 2\nu'\arctan\tau}\frac{I_0(4\sqrt{\rho\nu}\tau)}{\sqrt{1+\tau^2}},
    \end{equation}
    where $I_0$ is the modified Bessel function of the first kind. Detailed derivation can be found in the Appendix.
    
    According to \eqref{eq:decay_rate_via_imaginary_part}, one should then project an imaginary part of self energy onto the proton wave function, which is nontrivial due to the external field. We do not carry out this projection explicitly. Instead, it is natural to expect that the leading decay exponent is given only by an imaginary part of self energy, so we will only examine the asymptotics of $\Im\Sigma$ itself.

    Considering the weak field limit $\nu,\nu',\rho \gg 1$, we can use Bessel function asymptotics $I_0(z)\sim \frac{e^z}{\sqrt{2\pi z}}$, so leaving only the exponential order, we have
    \begin{equation}
        \mathcal I(k_\perp) :=\frac{d\Im\Sigma}{d^2k_\perp} \sim \int_0^\infty d\tau e^{-S(\tau)},
    \end{equation}
    where
    \begin{equation}
        S(\tau) = 2\pi\nu' + 2(\rho+\nu)\tau - 2\nu'\arctan\tau - 4\sqrt{\rho\nu}\tau.
    \end{equation}
    Denoting $\Delta = |\sqrt\nu-\sqrt\rho|$, we have the following.
    \begin{equation}
        S(\tau) = 2\pi\nu' + 2\Delta^2\tau - 2\nu'\arctan\tau.
    \end{equation}
    In the case $\Delta^2 < \nu'$ (which is actually the case for the decay process at hand), there is a saddle point $\tau_* = \sqrt{\frac{\nu'}{\Delta^2} - 1}$, and therefore
    \begin{equation}
        \mathcal I(k_\perp) \sim e^{-S_*},\quad S_*=S(\tau_*) = \pi\nu'+2\Delta\sqrt{\nu'-\Delta^2} + 2\nu'\arcsin\frac{\Delta}{\sqrt{\nu'}}.
    \end{equation}
    For the physically relevant hierarchy $\nu' \ll \nu,\rho$ and $\Delta^2 \ll \nu'$, we obtain
    \begin{equation}
        S_* = \pi\nu' + 4\Delta\sqrt{\nu'} + \mathcal O(\frac{\Delta^3}{\sqrt{\nu'}}).
    \end{equation}
    The complete integral $\Im\Sigma \sim \int dk\ ke^{-S_*(k)}$ builds up in the small neighborhood of $k=0$, and hence
    \begin{equation}
        \Im\Sigma \sim e^{-\frac{\pi m_\pi^2}{2|eE|}-\frac{2m_\pi}{|eE|}(m_n-m_p)},
    \end{equation}
    which reproduces the weak-field asymptotics of the instanton exponent.

\section{Entanglement}
    In this Section we comment on the entanglement between daughter particles in the post-selected two-body decay state. Since the daughter particles carry opposite transverse momenta, the decay produces a momentum-entangled state. It is useful to introduce momentum resolution $\Delta k$, or equivalently  the coherence scale $R_\text{coh}\sim\Delta k^{-1}$.
    
    Our Hilbert space is decomposed as
    \begin{equation}
        \mathcal H = \mathcal H_n \otimes \mathcal H_\pi, \qquad \mathcal H_n = \text{span}\{|n(k_\perp)\rangle\},\quad H_\pi = \text{span}\{|\pi(k_\perp)\rangle\}.
    \end{equation}
    Therefore, the wavefunction has the form
    \begin{equation}
        |\Psi\rangle = \int d^2k_\perp \sqrt{w(k_\perp)}e^{i\phi(k_\perp)}|n(k_\perp)\rangle\otimes |\pi(-k_\perp)\rangle,
    \end{equation}
    where $w$ is a momentum-space probability distribution, and is identified with normalized differential decay probability:
    \begin{equation}
        w(k_\perp) = \frac{1}{\Gamma}\frac{d\Gamma(k_\perp)}{d^2k_\perp},\qquad \int d^2 k_\perp w(k_\perp) = 1.
    \end{equation}
    Using orthogonality $\langle \pi(-k')|\pi(-k)\rangle \sim \delta^{(2)}(k-k')$, we obtain the  density matrix of neutral particle:
    \begin{equation}
        \rho_n = \int d^2k_\perp w(k_\perp)|n(k_\perp)\rangle\langle n(k_\perp)|.
    \end{equation}

     The discrete probability is
    \begin{equation}
        w_{mn} = \int_{\text{cell}(m,n)} d^2k\ w(k) \approx w(k_{mn})(\Delta k)^2,\qquad \sum_{m,n}w_{mn} = 1. 
    \end{equation}
   yielding Von Neumann entropy in discretized case :
    \begin{equation}
        \begin{gathered}
            S = -\tr \rho_n \ln\rho_n = -\sum_{m,n}w_{mn}\ln(w_{mn}) = \\ =-\sum_{m,n}w(k_{mn})(\Delta k)^2\ln\big[w(k_{mn})(\Delta k)^2)\big].
        \end{gathered}
    \end{equation}
    or in continuum limit ,
    \begin{equation}
        S = -\int d^2k\ w(k)\ln w(k) - \ln(\Delta k)^2.
    \end{equation}
    Note that if one naively introduced entropy only with a first term, it will not be invariant under coordinate rescaling $k=\lambda q$, $\lambda > 0$.  Now, substituting the decay rate, we find
    \begin{equation}
        w(k) = \frac{1}{\Gamma}\frac{d\Gamma}{d^2k} \sim e^{-\frac{\pi}{2|eE|}k^2} = e^{-\alpha k^2},\qquad \alpha = \frac{\pi}{2|eE|}.
    \end{equation}
    The normalization condition fixes the prefactor:
    \begin{equation}
        w(k) = \frac{\alpha}{\pi}e^{-\alpha k^2}.
    \end{equation}
    Note that this answer is obtained with logarithmic accuracy, as we ignored all dependence of the prefactor of $d\Gamma(k)/d^2k$ on k. Therefore, for entropy one obtains
    \begin{equation}
        S_\text{ent} = 1 + \ln\left(\frac{\pi}{\alpha(\Delta k)^2}\right) = 1 + \ln\left(\frac{2|eE|}{(\Delta k)^2}\right).
    \end{equation}

    Now we should comment on the value of $\Delta k$. It is set by the coherence length $R_\text{coh}$ as $\Delta k \sim \frac{1}{R_\text{coh}}$. There are several convenient ways to choose this length. Firstly, if the theory is placed in a box with size $L$, for example, the size of the region over which the electric field may be treated as constant. Then $R_\text{coh} = L$, and $\Delta k \sim \frac{1}{L}$. Secondly, this length can be determined by the size of the interaction vertex. 
    
\section{ Weak $p\rightarrow ne^{+}\nu_e$  decay in electric field}

    Let us turn to the decay of the charged particle with three particles in the final state, one charged, and two neutral. This mimics the  proton decay
    with the weak vertex substituting the strong interaction vertex considered in the previous sections. The energy that must be taken from the electric field in this case is essentially smaller since $m_n+m_e -m_p$ is smaller than $m_n+m_{\pi} -m_p$. The proton decay in the plane wave background  in this channel has been discussed in  \cite{vanzella2000weak,vanzella2001decay,wistisen2021transmutation}. The decay rate has been evaluated there by substituting the exact wave functions of the proton and positron in the plane wave into the weak $pne^+\nu_e$ vertex. 

\begin{figure}[h!]
    	\centering
    	\begin{tikzpicture}[thick,scale=1]
            \coordinate (T) at (0,1);
            \coordinate (B) at (0,-1);
            \coordinate (LT) at (-1.25,2);
            \coordinate (LB) at (-1.25,-2);

            \draw[midarrow] (B) to node[midway,left=2pt]{$n,\nu_e$} (T);
            \draw[midarrow] (LB) to [out=10,in=-110] node[pos=0.7, left=4pt] {$p$} (B);
            \draw[midarrow] (T) to [out=110,in=-10] node[pos=0.3,left=4pt] {$p$} (LT);

            \draw[midarrow] (B) arc[start angle=-110, end angle=110, radius=1.05] node[midway,right=4pt] {$e$};
            
        \end{tikzpicture}
    	\caption{Bounce for weak proton decay in Euclidean coordinates}
    	\label{fig:weak_bounce}
    \end{figure}
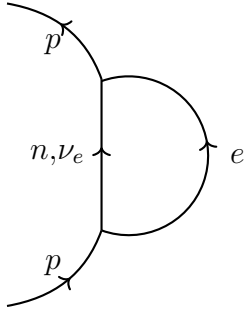

    Let us describe the proton decay rate involving the weak vertex via the worldline instanton configuration. It is important here that, even though the neutrino is a very light particle, it is still massive. The bounce configuration that involves massless particles is completely different and is not discussed here. Given that euclidean action contains two neutral contributions:
    \begin{equation}
        S_\text{neutral} = m_n L_n + m_\nu L_\nu.
    \end{equation}
    The Euclidean equation of motion for both neutron and neutrino is identical, and both worldlines connect the same junction points. Hence, both particles follow the same neutral line in the composite instanton (Fig. \ref{fig:weak_bounce}).
    
    Therefore:
    \begin{equation}
        L_n = L_\nu = L_N,\qquad u_n = u_\nu = u_N,
    \end{equation}
    so the neutral action contribution and the corresponding momentum conservation equation become
    \begin{equation}
        S_\text{neutral} = (m_n + m_\nu)L_N,\qquad m_pu_p = (m_n+m_\nu)u_N + m_eu_e.
    \end{equation}
    That is, the three-body problem at hand is reduced to the two-body problem, equivalent to the one considered in the previous sections. Therefore, we can write the instanton action in the very same form as \eqref{eq:instanton_action}, with adjusted masses $m_n \rightarrow  m_n + m_\nu$, $m_\pi \rightarrow m_e$. Given that $m_\nu \ll m_n$, the neutrino mass can be neglected and we have
    \begin{equation} \label{eq:magnetic_instanton_action}
        \begin{gathered}
            S_\text{inst} = \frac{m_e^2}{eE}\arccos\left(\frac{m_p^2 - m_e^2 - m_n^2}{2m_e m_n}\right) - \frac{m_p^2}{eE}\arccos\left(\frac{m_p^2 + m_n^2 - m_e^2}{2m_p m_n}\right) +
            \\ + \frac{m_e m_n}{eE}\sqrt{1-\left(\frac{m_e^2 + m_n^2 - m_p^2}{2m_e m_n}\right)^2}.
        \end{gathered}
    \end{equation}
    One more significant difference from the decay case $p\rightarrow n\pi$ is the fact that $m_e < m_n - m_p$. So, according to the discussions at the end of section \ref{sec:cartesian_coordinates}, the worldline instanton is complex, and the real part of the action is
    \begin{equation}
        \Re S_\text{inst} = \frac{\pi m_e^2}{|eE|}.
    \end{equation}
    We must once again highlight that this reduction to the two-body problem holds only at the level of instanton action, that is, at the leading exponential accuracy. The three-body nature of the problem is still contained within the prefactor.

    Next, we give some comments on the entanglement. The most natural analog of entanglement considered in the previous section is the bipartite entanglement between the charged positron and the neutral cluster $n\nu_e$. This is so because at the leading exponent accuracy, the instanton resolves only the total transverse recoil momentum of the neutral cluster $K_\perp = k_{n,\perp} + k_{\nu,\perp}$, while the partition of the momentum between the neutron and neutrino is not determined. For the rest, the calculation of the entanglement entropy is completely similar to that described in the previous section and leads to a similar answer:
    \begin{equation}
        S_{e:(n\nu)} = 1 + \ln\left(\frac{2|eE|}{(\Delta k)^2}\right).
    \end{equation}
    The full three-body entanglement would require the full differential probability beyond leading exponential accuracy, and hence is not discussed here.

\section{Charged particle decay in magnetic field}
    Let us now consider charged particle decay in an external magnetic field. We will first provide the calculation based on the overlap of wave functions and then show how to obtain the 
same result semiclassically. 
    
    \subsection{Wavefunctions overlap}
    Consider the wave function of a charged particle in an external magnetic field in Landau gauge : $H\parallel \hat x,A_2=0, A_3=Hx_2$, and denote $b = eH$. Then, the Hamiltonian is 
    \begin{equation}
        H = \sqrt{m^2 + p_2^2 + (p_3 - bx_2)^2}.
    \end{equation}
    The physical momenta operators are
    \begin{equation}
        \Pi_2 = p_2 \rightarrow -i\partial_2,\qquad \Pi_3 = p_3 - eA_3 \rightarrow -i\partial_3 - bx_2.
    \end{equation}
    Note that $x_3$ is cyclic and therefore the wavefunction is simply $\psi(x_2,x_3) = e^{iax_3}f(x_2)$, and $\Pi_3\phi = (a-bx)\psi$.

    In a magnetic field, wave functions are energy eigenstates corresponding to Landau levels: $\Pi_\perp^2\psi = P^2\psi$, where $P^2 = (2N+1)b$, $N$ --- number of Landau level. Substituting $\Pi_\perp^2 = \Pi_2^2 + \Pi_3^2$, we obtain an equation on $f(x_2)$:
    \begin{equation}
        \left[ -\frac{d^2}{dx^2} + b^2(x_2 - \frac{a}{b})^2\right] f(x_2) = (2N+1)bf(x_2).
    \end{equation}
    This is a rescaled and shifted equation of the harmonic oscillator and therefore has exact solution:
    \begin{equation}
        f_N(x_2) = \phi_N(x_2 - \frac{a}{b}),\qquad \phi_N(x) = \frac{b^{1/4}}{\pi^{1/4}\sqrt{2^N N!}}H_{N}(\sqrt{b}x)e^{-\frac{b}{2}x^2},
    \end{equation}
    where $H_N(x)$ is the Hermite polynomial: $H_N(x) = (-1)^Ne^{x^2}\frac{d^N}{dx^N}e^{-x^2}$.
    
    Therefore, the full expression for charged particle wavefunction in a magnetic field is 
    \begin{equation}
        u_{N,p_1,a} = \frac{1}{2\pi}e^{ip_1x_1}e^{iax_3}\phi_N(x_2 - \frac{a}{b}),
    \end{equation}
    and for a neutral particle the wave function is a plane  wave:
    \begin{equation}
        v_{\vec k}(\vec x) = \frac{1}{(2\pi)^{3/2}}e^{i\vec k\vec x}.
    \end{equation}

    In the stationary case, where $\Psi(x) = e^{-iEt}\psi(\vec x)$, the decay rate is driven by the spatial overlap of the wave functions. Setting $p_{1,p} = 0$ for simplicity, we have:
    \begin{equation}
        M_{p\rightarrow nc} = g\int d^3xu_{N_c,p_{1c},a_c}^*(\vec x)v_{\vec k}^*(\vec x)u_{N_p,0,a_p}(\vec x).
    \end{equation}
    Substituting wavefunctions and performing $x_1$ and $x_3$ integrals, we obtain, up to the constant factors :
    \begin{equation}
        M_{p\rightarrow nc}\propto I_{N_pN_c}(k_2,k_3) = \int dx e^{-ik_2x}\phi_{N_c}(x + \frac{k_3}{b})\phi_{N_p}(x),
    \end{equation}
    and up to the prefactors the decay rate is given by
    \begin{equation}
        \Gamma \propto \sum_{N_c}\int dk_1 dk_2 dk_3 |I_{N_pN_c}(k_2,k_3)|^2\delta(E_p - E_c - E_n).
    \end{equation}

    Now our goal is to derive the expression for $I_{N_pN_c}$. Note that $\phi_N(x+s) = e^{isP}\phi_N(x)$, where $P=-i\frac{d}{dx}$. Then, $I_{N_pN_c}$ can be written as matrix element $\bra{N_c}e^{-i\frac{k_3}{b}P}e^{-ik_2x}\ket{N_p}$. As $\ket N$ is the harmonic oscillator state, we can introduce ladder operators as $x = \frac{1}{\sqrt{2b}}(a+a^\dagger), P = -i\sqrt{\frac{b}{2}}(a - a^\dagger)$, so the matrix element can be reduced to the following form:
    \begin{equation}
        I_{N_p N_c} = e^{\frac{ik_2 k_3}{2b}}\bra{N_c}e^{-\frac{|\alpha|^2}{2}e^{\alpha a^\dagger}e^{-\alpha^*a}}\ket{N_p}, \qquad \text{where } \alpha = \frac{k_3 - ik_2}{\sqrt{2b}}.
    \end{equation}
    This expression can be easily calculated by how ladder operators act in states $\ket N$, and the answer for the integrand inside the decay rate is
    \begin{equation}
        |I_{N_pN_c}(k_2,k_3)|^2 = e^{-\frac{K^2}{2b}}\frac{n!}{(n+l)!}\left(\frac{K^2}{2b}\right)^l \Big[L_n^l \left(\frac{K^2}{2b}\right)\Big]^2,
    \end{equation}
    where $L_n^k(u) = \sum_{s=0}^n \frac{(-1)^s (n+k)!}{(n-s)!(k+s)!s!}u^s$ are associated Laguerre polynomials, and the notation $K^2 = k_2^2 + k_3^2$, $n = \min(N_p,N_c)$ and $l = |N_p - N_c|$ is introduced.

    Now we will simplify this expression in the limit of a weak field. Indeed, set $b\rightarrow 0$, $N_p,N_c\rightarrow \infty$, while keeping $P_p^2 = (2N_p+1)b$ and $P_c^2 = (2N_c + 1)b$ fixed. Assume $P_p > P_c$, as this is the case for the decay process, and consider the following notation, where we drop the $+1$ term inside the definition of Landau levels due to $N_p,N_c \gg 1$:
    \begin{equation}
        \begin{gathered}
        N_c = \frac{A}{b}, \quad A=\frac{P_c^2}{2},\\
        N_p = \frac{Q}{b},\quad Q = \frac{P_p^2}{2},\\
        N_p - N_c = \frac{B}{b}, \quad B = Q-A = \frac{P_p^2 - P_c^2}{2},\\
        C = \frac{K^2}{2}.
        \end{gathered}
    \end{equation}
    Now consider the integral representation of associated Lagger polynomials:
    \begin{equation}
        L_n^k(x) = \frac{1}{2\pi i}\int_\gamma \frac{e^{-\frac{xz}{1-z}}}{(1-z)^{k+1}z^{n+1}}dz,
    \end{equation}
    where the contour $\gamma$ is around $z=0$ and does not contain $z=1$. For our case, we have the following.
    \begin{equation}
        L_{A/b}^{B/b}\left(\frac{C}{b}\right) \propto \int_\gamma dz e^{\frac{1}{b}f(z)},\qquad f(z) = -A\log z - B\log (1-z) - C\frac{z}{1-z},
    \end{equation}
    where the prefactor was dropped and also we ignored some terms due to $b\rightarrow 0$. This contour integral can be calculated via the method of stationary phase:
    \begin{equation}
        f'(z) = 0 \Rightarrow z_{\pm} = \frac{A+Q-C \pm \sqrt{(A+Q-C)^2-4AQ}}{2Q}.
    \end{equation}
    The minimum of $f(x)$ is achieved at $z_* = z_{-}$, and the integral can be approximated as $L_{A/b}^{B/b}\left(\frac{C}{b}\right) \sim e^{\frac{1}{b}f(z_*)}$. 

    Using the Stirling formula, we obtain the following result.
    \begin{equation}
        |I_{N_pN_c}|^2 = e^{-\frac{1}{b}S},\quad S = C - B + A\log\frac{Q}{A} + B\log\frac{Q}{B} + 2A\log z_* + 2B\log(1-z_*) + \frac{2Cz_*}{1-z_*}.
    \end{equation}
    Let us introduce the following.
    \begin{equation}
        Y = P_p\sinh\alpha_p = P_c\sinh\alpha_c,\qquad K = P_p\cosh\alpha_p - P_c\cosh\alpha_c.
    \end{equation}
    This parametrization is valid for $P_p > P_c$ and $0<K<P_p - P_c$, which is precisely the physical decay considered here. By straightforward calculation, it can be shown that the "action" $S$ is reduced to a simple  expression
    \begin{equation}
        S = P_p^2 \alpha_p - P_c^2\alpha_c - KY.
    \end{equation}

    Substitute this result into the decay rate:
    \begin{equation}
        \Gamma \propto \int dP_c^2 dk_1 dK^2\ \exp\left(-\frac{P_p^2\alpha_p - P_c^2\alpha_c - KY}{eH}\right)\delta(E_p - E_c - E_n),
    \end{equation}
    where we switched from summation to integration $\sum_{N_c} \rightarrow \int dP_c^2$, as at the weak field limit Landau levels are almost continuous. The $K^2$ integration is performed using the energy-conservation delta function, which imposes the constraint
    \begin{equation}
        K^2 = \Bigg[E_p - \sqrt{m_c^2 + P_c^2 + k_1^2}\Bigg]^2 - m_n^2 - k_1^2.
    \end{equation}
    The saddle with respect to $k_1$ is $k_1 = 0$. Finally, taking the saddle with respect to $P_c$ we obtain
    \begin{equation}
        \Gamma \sim \exp\left[-\frac{S_*}{eH}\right], \qquad S_* = \min_{P_c} S(P_p,P_c, K(P_p, P_c)).
    \end{equation}
    This saddle gives the equation
    \begin{equation}
        \frac{dS}{dP_c} = \frac{\partial S}{\partial P_c} + \frac{\partial S}{\partial K}\frac{dK}{dP_c} = 0 \Rightarrow \alpha_c = \frac{YE_n}{KE_c}.
    \end{equation}
    Thus, the decay rate can be found from the system
    \begin{equation} \label{eq:magnetic_system}
        \begin{cases}
            Y = P_p\sinh\alpha_p = P_c\sinh\alpha_c,\\
            K = P_p\cosh\alpha_p - P_c\cosh\alpha_c,\\
            E_p = E_c + E_n,\\
            E_i^2 = m_i^2 + P_i^2,\quad i\in\{p,c,n\},\\
            \alpha_c = \frac{YE_n}{KE_c},\\
            \Gamma \sim \exp\left[-\frac{P_p^2\alpha_p - P_c^2\alpha_c - KY}{eH}\right].
        \end{cases}
    \end{equation}
    This system cannot be solved analytically, so we will consider the two most interesting limit cases. Firstly, we will consider the limit of the initial ultrarelativistic(UR) proton. Then we will consider the reverse case, where the proton energy is barely enough for decay.

    \subsection{Ultrarelativistic limit}
    Let us solve the system \eqref{eq:magnetic_system} in the limit of the ultrarelativistic incoming proton. Let $E_p = E$, $E\gg m_p,m_n,m_c$, and $E_c = xE$, $E_n = (1-x)E$ for some $0<x<1$. Then
    \begin{equation}
        P_p = E - \frac{m_p^2}{2E}, \qquad P_c = xE - \frac{m_c^2}{2xE}, \qquad K = (1-x)E - \frac{m_n^2}{2(1-x)E}.
    \end{equation} up to some order of accuracy. It can be shown that $Y$ is $O(1)$ in UR limit. Hence, for $\alpha_c$ we have
    \begin{equation}
        \alpha_c = \frac{Y}{xE} + \frac{1}{E^3}\left(\frac{Ym_c^2}{2x^3} - \frac{Y^3}{6x^3}\right)
    \end{equation}
    up to the same order of accuracy. Given that, the saddle equation of $\alpha_c$ gives
    \begin{equation}
        Y^2 = 3\left(m_c^2 - \frac{x^2m_n^2}{(1-x)^2}\right).
    \end{equation}
    hence substituting this expression into $K = P_p\cosh\alpha_p - P_c\cosh\alpha_c$ we obtain the quadratic equation on $x$:
    \begin{equation}
        (m_p^2 + 3m_n^2 - 3m_c^2)x^2 + (m_n^2 + 5m_c^2 - m_p^2)x -2m_c^2 = 0.
    \end{equation}
    Its  solution
    \begin{equation}
        x_* = \frac{m_p^2 - m_n^2 - 5m_c^2 + \sqrt{(m_n^2 + 5m_c^2 - m_p^2)^2 + 8m_c^2(m_p^2 + 3m_n^2 - 3m_c^2)}}{2(m_p^2 + 3m_n^2 - 3m_c^2)},
    \end{equation}
    provides the UR decay exponent; 
    \begin{equation} \label{eq:UR_decay_action_x}
        S^\text{UR} = \frac{2\sqrt 3 M}{eHE_p}\frac{1-x_*}{x_*}\left(m_c^2 - \frac{x_*^2m_n^2}{(1-x_*)^2}\right)^{3/2}.
    \end{equation}

    Consider the following physical mass hierarchy: $m_p = M, m_n = M+\delta, m_c = \mu$ and $\delta,\mu \ll M$. Then expressions are hugely simplified and lead to
    \begin{equation} \label{eq:UR_decay_action}
        S^\text{UR} = \frac{\sqrt 3}{8eHE_p}\sqrt{s-\delta}(s+3\delta)^{3/2}, \qquad s = \sqrt{\delta^2 + 8\mu^2}.
    \end{equation}

    Now we will comment on the case of a decay in a crossed field. Consider the UR proton in a pure magnetic field. Perform a boost perpendicular to the field direction with some velocity $\beta$. Then the EM field components are $\vec E' = \gamma \vec\beta \times\vec B$, $\vec B' = \gamma \vec B$, so the components are perpendicular to each other, and the relative difference between them is $\frac{|\vec B'| - |\vec E'|}{|\vec B'|} = 1-\beta \approx \frac{1}{2\gamma^2}$. This leads us to the fact that the UR particle in a pure magnetic field is equivalent to a soft particle in an almost crossed field. This suggests assigning the decay exponent \eqref{eq:UR_decay_action} to the decay of a soft proton in a crossed field and vice versa.
    An answer for the decay of the proton in the crossed field was obtained in \cite{wistisen2021transmutation} through the Volkov states. By direct substitution, we can verify that this answer matches \eqref{eq:UR_decay_action}, which provides another independent check of the obtained result.

    \subsection{Soft proton limit}
    Now let us turn to solving the system \eqref{eq:magnetic_system} in the limit, where the proton energy is barely enough for the decay. This is imposed by letting
    \begin{equation}
        E = E_p,\quad P = P_p = \sqrt{E^2 - m_p^2}, \quad E = m_n + m_c + \ve,\quad 0<\ve\ll m_n,m_c.
    \end{equation}
    Then it is clear that $P_c^2 = O(\ve)$, $K^2 = O(\ve)$. Moreover, from saddle equations one finds that $P_c$ is exponentially suppressed in $\ve$. Therefore, it can be set to zero during the calculation, and the saddle equations become just that
    \begin{equation}
        Y = P\sinh\alpha_p,\qquad K = P\cosh\alpha_p - Y,
    \end{equation}
    with a solution
    \begin{equation}
        Y = \frac{P^2 - K^2}{2K},\qquad \alpha_p = \log\frac{P}{K}.
    \end{equation}
    Substituting this into the expression for $S$ and introducing the notation $P_0^2 = (m_n + m_c)^2 - m_p^2$, we obtain
    \begin{equation}
        S^\text{soft} = \frac{P_0^2}{2eH}\log\frac{1}{\ve} + O(1).
    \end{equation}
    Thus, we got the expected result: the decay exponent explodes as the proton energy approaches the threshold energy: $S^\text{soft} \rightarrow +\infty$ as $\ve \rightarrow +0$. This is strong evidence that there is no decay below the threshold energy, as one would normally assume from the energy conservation principle.

    \subsection{Magnetic instanton}
    Now we will show how the result \eqref{eq:magnetic_system} can be obtained semiclassically, using the composite worldline instanton. Naively, one may think of considering the same euclidean action as we used in the electric decay case. However, this turns out to be incorrect. The subtle point is that in magnetic decay the tunneling effectively occurs not between two spacetime points, but between two Landau level states. Informally speaking, here we have tunneling with prescribed energy and momentum, not coordinate (which was the case for electric tunneling, where spacetime shift between turning points was fixed by the amount of energy gained from the field to overcome the energy gap). 

    For the neutral initial particle, the exponentially suppressed decay of a photon or neutrino in the magnetic field has been discussed in \cite{satunin2013width, satunin2015study} in the corresponding range of parameters. The composite worldline instanton was found to occur in complexified coordinates, being an example of the complex instanton somewhat similar to the complex instantons in time-dependent electric fields \cite{popov1972pair,Dumlu_2011}. The effective action once again involves the length and area contributions, but the composite curve is embedded in a three-dimensional space. The presence of a negative mode in the spectrum of fluctuations is confirmed.

    We follow a bit different route for the magnetic case and argue 
    that the result for the decay probability obtained by the overlap of the wave functions is reproduced if the Routhian action is used.
    Therefore, consider the Routhian transform of the standard proper-time action with respect to energy and conserved momentum along the $y$ axis:
    \begin{equation} \label{eq:Routh_transform}
        \tilde S = S - p_y\Delta y + E\Delta t,
    \end{equation}
    where
    \begin{equation}
        S = -m\int ds + b\int xdy,
    \end{equation}
    and $A_x=0,A_y=Hx$ , $b=eH$. Therefore, the Routhian action is
    \begin{equation}
        \tilde S = \int p_xdx = \int dx\sqrt{P^2 - (p_y - bx)^2}.
    \end{equation}
    Now consider the tunneling region, where the momentum is imaginary. Upon Wick rotation  $p_x = i\pi$ and $\tilde S_E = -i\tilde S = \int pdx$, where the condition in the shell is $\pi^2 - \Pi_y^2 - P^2$. The equation of motions can be obtained if one restores the on-shell constraint inside the action as
    \begin{equation}
        \tilde S_E = \int d\lambda\left[\pi\dot x - \frac{N}{2}(\pi^2 - \Pi_y^2 + P^2)\right].
    \end{equation}
    Varying this action and setting convenient parameterizations such as $N = \frac{1}{E}$, we obtain
    \begin{equation}
        \dot x = \frac{\pi}{E},\quad \dot\pi = -\frac{b}{E}\Pi_y, \quad \dot \Pi_y = -\frac{b}{E}\pi.
    \end{equation}
    The solution can be written in the following parametrization:
    \begin{equation}
        x = \frac{p_y - P\cosh\alpha}{b}, \quad \pi = -P\sinh\alpha, \quad \Pi_y = P\cosh\alpha.
    \end{equation}
    
    The action evaluated on these trajectories is
    \begin{equation}
        \tilde S_E = \frac{P^2}{b}\int_{-\alpha}^\alpha \sinh^2\tilde\alpha d\tilde\alpha = \frac{P^2}{b}(\sinh\alpha\cosh\alpha - \alpha),
    \end{equation}
    where we used the symmetry of the bounce. The action on the neutral trajectory is the following.
    \begin{equation}
        \tilde S_{E,n} = \int \pi_ndx_n.
    \end{equation}
    For neutral particles, $\pi_n = \text{const}$ and $\tilde S_{E,n} = \pi_n\Delta x_n$. Varying with respect to $\Delta x_n$, we obtain $\pi_n = 0$, $\tilde S_{E,n} = 0$.

    The decay rate is given by the composite instanton, whose action is
    \begin{equation}
        \tilde S_\text{inst} = \tilde S_{E,c} + \tilde S_{E,n} - \tilde S_{E,p} = \frac{1}{b}\Big[ P_c^2(\sinh\alpha_c\cosh\alpha_c - \alpha_c) - P_p^2(\sinh\alpha_p\cosh\alpha_p - \alpha_p) \Big].
    \end{equation}
    The momentum conservation law at the vertices is expressed as
    \begin{equation}
        Y = P_p\sinh\alpha_p = P_c\sinh\alpha_c,\qquad K = P_p\cosh\alpha_p - P_c\cosh\alpha_c,
    \end{equation}
    and in these terms the instanton action is
    \begin{equation}
        \tilde S_\text{inst} = \frac{P_p^2\alpha_p - P_c^2\alpha_c - KY}{eH}.
    \end{equation}
    Varying this with respect to $P_c$ gives the required equation in $\alpha_c$, and thus we have obtained the  same system \eqref{eq:magnetic_system} as derived from the overlap approach.

    \subsection{Generalized synchrotron emission}
    The process in question can be viewed as a generalization of standard synchrotron emission. Indeed, the situation is very similar: we have a particle on some Landau level that emits another particle, carrying away part of the momentum and energy, and therefore the initial particle jumps to some lower Landau level. The decay rate is given by the overlap of these two Landau levels. The main distinction between standard synchrotron emission and the process discussed in this work is that the former is allowed kinematically. Exactly, synchrotron emission is driven by very soft photons, so the process is a combination of many processes 
    \begin{equation}
        e^-\rightarrow e^- +\gamma
    \end{equation}
    with a vanishing mass gap. So, the initial and final electrons sit on almost the  same Landau levels, whose overlap is almost not suppressed. This leads to the fact that the total rate for synchrotron emission is power-like, not exponentially suppressed because it is driven by the exponent prefactor, not the suppressing exponent.

    However, the differential rate for hard photons is exponentially suppressed in agreement with \eqref{eq:magnetic_system}. We will show this explicitly in UR limit. The standard ultrarelativistic semiclassical quantum-synchrotron formula is well known and can be found, for example, in \cite{blackburn2020}:
    \begin{equation}
        \frac{d\Gamma_\text{syn}}{d\omega} = \frac{\alpha}{\sqrt{3}\pi\gamma^2}\Bigg[ \left(x + \frac{1}{x}\right)K_{2/3}(\xi) - \int_\xi^\infty K_{1/3}(y)dy \Bigg],
    \end{equation}
    where $\gamma = \frac{E}{m}, x = \frac{E - \omega}{E}, \xi = \frac{2(1-x)}{3\chi}$, and for the pure magnetic field $\chi = \frac{bP}{m^3}$. In the limit of weak field $\chi\rightarrow0, \xi\rightarrow+\infty$, so expanding Bessel functions
    \begin{equation}
        K_\nu(\xi) = \sqrt{\frac{\pi}{2\xi}}e^{-\xi}\left(1 + \frac{4\nu^2 - 1}{8\xi} + O(\xi^{-2})\right),\quad \xi\rightarrow \infty,
    \end{equation}
    we obtain
    \begin{equation} \label{eq:synchrotron_rate_standard}
        \frac{d\Gamma_\text{syn}}{d\omega} = \frac{\alpha}{\sqrt{3}\pi\gamma^2}\sqrt{\frac{\pi}{2\xi}}e^{-\xi}\left(\frac{1-x+x^2}{x} + O(\xi^{-1})\right) \sim e^{-\xi} = \exp\left[-\frac{1}{eH}\frac{2m^3}{3P}\frac{1-x}{x}\right].
    \end{equation}

    Now we impose the synchrotron emission setup on our calculation. Exactly, set $m_p = m_c = m$, $m_n = 0$, and fix the photon energy $\omega$. This fixation prevents us from using the equation $\alpha_c = \frac{YE_n}{KE_c}$ from system \eqref{eq:magnetic_system}, since this equation was obtained by minimizing the action on $P_c$. As we fix $\omega$, $P_c$ is also fixed. So, usage of \eqref{eq:UR_decay_action_x} is not allowed. Instead, solving
    \begin{equation}
        Y = P_p\sinh\alpha_p = P_c\sinh\alpha_c,\qquad K = P_p\cosh\alpha_p - P_c\cosh\alpha_c,
    \end{equation}
    we obtain $Y=m$ and
    \begin{equation}
        eHS_\text{syn} = (E^2-m^2)\arctanh\frac{m}{E}-(x^2E^2 - m^2)\arctanh\frac{m}{xE} - (1-x)Em.
    \end{equation}
    Expanding this into the UR limit $E,xE\gg m$, we obtain
    \begin{equation}
        S_\text{syn}^\text{UR} = \frac{1}{eH}\frac{2m^3}{3P}\frac{1-x}{x},\qquad \frac{d\Gamma}{d\omega} = e^{-S_\text{syn}^\text{UR}},
    \end{equation}
    which coincides with \eqref{eq:synchrotron_rate_standard}.

    If, instead, we let $\omega$ vary freely, we will obtain $S \rightarrow 0$ at $\omega\rightarrow 0$, which corresponds to a power-like rate driven by soft photons, as was discussed at the beginning of the section.

    \subsection{Induced decay rate}
    Now we shall discuss the photon-assisted decay rate, i.e. the process
    \begin{equation} \label{eq:assisted_process}
        \gamma(q) + p \rightarrow c + n,\qquad q = (\Omega,\vec q).
    \end{equation}
    In contrast with induced Schwinger effect, the incoming photon assists not by effectively lowering the potential barrier, as we do not have one in our problem, but by modifying the momentum conservation law at the vertex in a way that brings the momenta closer to the kinematically allowed region. For the process \eqref{eq:assisted_process} the neutral wavefunction is
    \begin{equation}
        e^{-i(\vec k_\perp - \vec q_\perp)\cdot \vec x},
    \end{equation}
    Therefore, maximum assistance is obtained when $q_\parallel = 0$, $|q_\perp| = \Omega$ and the momentum is directed oppositely to the recoil of the neutral particle: $\vec q_\perp = -\vec k_\perp$, $K = |\vec k_\perp|$. Given that, we should replace $K\rightarrow K + \Omega$ in our considerations. The corresponding system is
    \begin{equation}
        \begin{cases}
            Y = P_p\sinh\alpha_p = P_c\sinh\alpha_c,\\
            K + \Omega = P_p\cosh\alpha_p - P_c\cosh\alpha_c,\\
            E_p + \Omega = E_c + E_n,\\
            E_i^2 = m_i^2 + P_i^2,\quad i\in\{p,c,n\},\\
            \alpha_c = \frac{YE_n}{KE_c},\\
            \Gamma \sim \exp\left[\frac{P_p^2\alpha_p - P_c^2\alpha_c - (K+\Omega)Y}{eH}\right].
        \end{cases}
    \end{equation}
    Note that since $\Omega$ is fixed and independent of $P_c$, the equation $\alpha_c = \frac{YE_n}{kE_c}$ is unchanged.

    Solving this system in the ultrarelativistic limit, we obtain a modified quadratic equation on $x = \frac{E_c}{E_p}$:
    \begin{equation} \label{eq:quadratic_equation_induced}
        (m_p^2 + 3m_n^2 - 3m_c^2 + 4\eta)x^2 + (m_n^2 + 5m_c^2 - m_p^2 - 4\eta)x - 2m_c^2 = 0,
    \end{equation}
    where $\eta = \Omega E_p$ was introduced, and for action we obtain the old expression
    \begin{equation}
        S_\eta^\text{UR} = \frac{2\sqrt{3}}{eHE_p}\frac{1-x_\eta}{x_\eta}\left(m_c^2 - \frac{x_\eta^2m_n^2}{(1-x_\eta)^2}\right)^{3/2},    
    \end{equation}
    where $x_\eta$ is a physical solution of the quadratic equation above.

    Here, one can derive the critical photon energy because the action vanishes at $x_\eta = \frac{m_c}{m_c + m_n}$. Substituting this into \eqref{eq:quadratic_equation_induced}, we find the critical value of the photon energy:
    \begin{equation}
        \Omega_\text{crit} = \frac{(m_c + m_n)^2 - m_p^2}{4E_p}.
    \end{equation}
    For $0 < \Omega < \Omega_\text{crit}$ the process is exponentially suppressed, and for $\Omega \geq \Omega_\text{crit}$ the process becomes kinematically allowed and has a power-like decay rate.

    Using the mass hierarchy $m_p = M, m_n = M+\delta, m_c = \mu$, $\delta, \mu \ll M$, we obtain
    \begin{equation}
        S_\eta^\text{UR} = \frac{\sqrt{3}M}{8eHE_p}\sqrt{s_\eta - \delta_\eta}(s_\eta + 3\delta_\eta)^3, \qquad s_\eta = \sqrt{\delta_\eta^2 + 8\mu^2},
    \end{equation}
    where $\delta_\eta = \delta - \frac{2E_p\Omega}{M}$. So, the incoming photon reduces the mass gap, completely wiping it out at critical energy $\Omega_\text{crit} = \frac{M(\delta + \mu)}{2E_p}$.

\section{Discussion}

In this study, we have examined the validity of the composite worldline instantons for the description of the
non-perturbative decay of the particles in constant external fields. Our key example is the nonperturbative 
decay of the charged particle in the external electric field. We have demonstrated explicitly the coincidence 
in the leading exponential approximation of the decay rate evaluated by the conventional imaginary part of the
self-energy and the composite worldline instanton. This supports the use of composite worldline instantons for an arbitrary number of particles in the final state. For charged-particle decay in a magnetic field, the wave-function-overlap result agrees, in the relevant limit, with the semiclassical calculation. Such a process can be considered as a generalized synchrotron emission.

The version of the statistical entanglement entropy of the created particles has been evaluated. It differs from 
the entanglement entropy discussed in \cite{ghodrati2015schwinger,grieninger2023entanglement} which is based on the thermodynamic analogy. It would be interesting to investigate possible thermodynamic versions of the entanglement entropy in the decay case similar to \cite{ghodrati2015schwinger,grieninger2023entanglement}.

Although our main goal was to test the composite worldline instanton approach, it is worth commenting on the applicability 
of the decay process to the real situations. In particular it was argued in \cite{dolgov2024conversion} that the
interesting process can be essential near the massive positively charged black holes in some range of parameters.
The magnitude of the electric field near the charged black hole was argued to be near critical for the Schwinger process \cite{dolgov2024conversion}.
The
protons are gravitationally attracted by the black hole, and simultaneously there is a Schwinger pair creation
near the horizon by an almost critical electric field. Hence, there is an effective conversion of the ingoing protons into 
the outgoing positrons. We note that at this parameter domain the proton decay near the horizon should be taken 
into account in the total conversion process. Indeed, the probability of decay of $p\rightarrow ne\nu$ in the
electric field is compared to the probability of Schwinger pair production. After the decay the neutron is falling
to the black hole and there is the positron outgoing flux. The emerging positron flux involves the soft  particles and 
is correlated with the neutrino flux.

There are a few questions for further research. The most evident question concerns the evaluation of the decay 
probability rate beyond the leading exponential approximation. To this aim, it is  necessary to take into account
the higher winding modes on the instanton configuration as well as the fluctuations of the 
trajectories involved in the composite instanton. It is interesting to generalize the composite 
worldline instantons for particle decay in inhomogeneous fields.

It would be interesting  to develop an analog of the $ER=EPR$
relation for particle decay. The key difference from the discussion in \cite{sonner2013holographic,jensen2013holographic} is that the natural bridge 
between the charged particles in the final state is now absent, hence there is no space-like entanglement entropy 
associated with the holographic string connecting the daughter particles. There 
is also no natural bridge between the initial and final charged particles, since they have the same sign charges.
However, formally the effective action for the decay process involves the closed Wilson loop in Euclidean space-time
built from the charged pion trajectory and the fictitious trajectory of "antiproton".
Therefore, we could make a fixed time slice of the minimal surface which has this fictitious Wilson loop as a boundary. Such a slice yields 
a kind of fictitious bridge whose meaning deserves interpretation.

Another natural question  concerns the strong coupling
probability rate for particles decays in external fields in holographic QCD or  equivalently  in chiral perturbation theory(ChPT). 
The instanton interpretation of baryon in holography 
is parallel to the Skyrmion interpretation in ChPT hence we can interpret the decay of baryon 
in an external field as instability of the  
instanton in holographic QCD or as discharge of the electrically charged Skyrmion through charged pion emission.
We hope to discuss this issue in a separate study.

The composite worldline or worldvolume instantons can also describe processes with the vacuum as the initial state and several charged and neutral particles in the final state. Such composite instantons were discussed in the
cosmological context in \cite{gorsky2000tunnelling,gorsky2007spontaneous}. It would be interesting to investigate such composite instantons in 
Schwinger-like processes and to investigate the domain of parameters when the thermal analogy will 
be applicable.

The authors thank P. Pakhlov for the useful discussions.

\section{Appendix}
    \subsection{Residue evaluation}
    For the calculation 
    \begin{equation}
        R(A,B) = \Res_{z=0}\frac{1}{\sin z}e^{A\cot z - Bz}.
    \end{equation}
    let us introduce $f(z) = \frac{1}{\sin z}e^{A\cot z - Bz}$. It is easy to see that this function satisfies the following differential equation
    \begin{equation}
        f_z + (\cot z + \frac{A}{\sin^2 z} + B)f = 0.
    \end{equation}
    Taking the residue of both sides of this equation and using $\Res_{z=0}(f_z) = 0$:
    \begin{equation}
        \Res_{z=0}(f\cot z) + A\Res_{z=0}(\frac{f}{\sin^2 z}) + B\Res_{z=0}f = 0.
    \end{equation}
    Note that $f\cot z = \partial_A f$ and $\frac{f}{\sin^2 z} = f + \partial_A^2 f$, therefore, we arrive at the following differential equation for $R$:
    \begin{equation} \label{eq:residue_DE}
        AR_{AA} + R_A + (A + B)R = 0.
    \end{equation}
    The initial condition at $A = 0$ is obtained from the coefficient of $z^{-1}$ in the Laurent series of $\frac{e^{-Bx}}{\sin z}$. Exactly, $R(0,B) = 1$. Considering ansatz $R(A,B) = e^{iA}Y(x,B)$, where $x = -2iA$, \eqref{eq:residue_DE} reduces to Kummer's differential equation
    \begin{equation}
        xY'' + (b-x)Y' - aY = 0
    \end{equation}
    with parameters $a = \frac{1-iB}{2},b=1$. The analytic solution at $x=0$, satisfying $Y(0) = 1$ is given by (Kummer's) confluent hypergeometric function:
    \begin{equation}
        Y(x) = {}_1F_1(a,b,x).
    \end{equation}
    Using Kummer's transformation
    \begin{equation}
        e^x{}_1F_1(a,1,-x) = {}_1F_1(1-a,1,x),
    \end{equation}
    we arrive at
    \begin{equation}
        R(A,B) = e^{-iA}{}_1F_1(\frac{1+iB}{2},1,2iA).
    \end{equation}

    \subsection{Self energy proper time integral derivation}
    To derive \eqref{eq:imaginary_part_final_integral} we start from the integrand of \eqref{eq:imaginary_sigma_intermediate} and first integrate over $s_n$:
    \begin{equation}
        \frac{d\Im\Sigma}{d^2k} = -\frac{\pi g^2}{a}\frac{e^{-\pi\nu'}}{1+e^{-2\pi\nu'}}\int\frac{d^2q}{(2\pi)^4}\frac{e^{-\frac{iq^2}{a}}{}_1F_1(\frac{1}{2} + i\nu', 1, \frac{2iq^2}{a})}{2a\rho + (p-q)^2}.
    \end{equation}
    In polar coordinates $d^2q = qdqd\phi$, $(p-q)^2 = p^2 + q^2 - 2pq\cos\phi$, and the angle integral is
    \begin{equation}
        \int_0^{2\pi}\frac{d\phi}{2a\rho + p^2 + q^2 - 2pq\cos\phi} = \frac{2\pi}{\sqrt{(2a\rho + p^2 + q^2)^2 - 4p^2q^2}}.
    \end{equation}
    Analytically continue to the proton mass shell $p_E^2 \mapsto -m_p^2 = -2a\nu$, and introduce dimensionless variable $x = \frac{q^2}{a}$:
    \begin{equation}
        \frac{d\Im\Sigma}{d^2k} = -\frac{g^2}{16\pi^2 a}\frac{e^{-\pi\nu'}}{1+e^{-2\pi\nu'}}\int_0^\infty dx\frac{e^{-ix}{}_1F_1(\frac{1}{2}+i\nu',1,2ix)}{\sqrt{(x + 2(\rho-\nu))^2 + 8\nu x}}.
    \end{equation}
    Noting that
    \begin{equation}
        \frac{1}{\sqrt{(x + 2(\rho-\nu))^2 + 8\nu x}} = \frac{1}{\sqrt{(x + 2(\rho+\nu))^2 - (4\sqrt{\rho\nu})^2}}
    \end{equation}
    and using standard identity for Bessel function
    \begin{equation}
        \int_0^\infty d\tau\ e^{-A\tau}I_0(B\tau) = \frac{1}{\sqrt{A^2 - B^2}} \qquad \text{for}\quad \Re A > |B|,
    \end{equation}
    we get
    \begin{equation}
        \begin{gathered}
            \frac{d\Im\Sigma}{d^2k} = -\frac{g^2}{16\pi^2 a}\frac{e^{-\pi\nu'}}{1+e^{-2\pi\nu'}}\int_0^\infty d\tau\ e^{-2(\rho + \nu)\tau}I_0(4\sqrt{\rho\nu}\tau) \times \\
            \times \int_0^\infty dx\ e^{-(\tau+i)x}{}_1F_1(\frac{1}{2}+i\nu',1,2ix),
        \end{gathered}
    \end{equation}
    where $I_0$ is the modified Bessel function of the first kind.

    Next, we use the Laplace transform of Kummer's function
    \begin{equation}
        \int_0^\infty dxe^{-sx}{}_1F_1(a,b,-x) = \frac{1}{s}{}_2F_1(1,a,b,-\frac{1}{s}),
    \end{equation}
    using ${}_2F_1(a,1,1,z) = (1-z)^{-a}$, for our situation, we obtain the following.
    \begin{equation}
        \begin{gathered}
            \int_0^\infty dx\ e^{-(\tau+i)x}{}_1F_1(a,1,2ix) = \frac{1}{\tau+i}{}_2F_1(a,1,1,\frac{2i}{\tau+i}) = \\
            = (\tau + i)^{a-1}(\tau-i)^{-a}.
        \end{gathered}
    \end{equation}
    Noting that $\tau\pm i = \sqrt{1+\tau^2}e^{\pm\arctan\frac{1}{\tau}}$, we have
    \begin{equation}
        (\tau + i)^{a-1}(\tau-i)^{-a} = \frac{e^{-\pi\nu'}}{\sqrt{1+\tau^2}}e^{2\nu'\arctan\tau}.
    \end{equation}
    Collecting these results, we obtain the desired expression
    \begin{equation}
        \frac{d\Im\Sigma}{d^2k} = -\frac{g^2}{16\pi^2 a}\frac{e^{-2\pi\nu'}}{1+e^{-2\pi\nu'}}\int_0^\infty d\tau\ e^{-2(\rho+\nu)\tau + 2\nu'\arctan\tau}\frac{I_0(4\sqrt{\rho\nu}\tau)}{\sqrt{1+\tau^2}}.
    \end{equation}

\bibliographystyle{unsrt}
\bibliography{ref.bib}

\begin{thebibliography}{10}

\bibitem{sauter1931verhalten}
Fritz Sauter.
\newblock {\"U}ber das verhalten eines elektrons im homogenen elektrischen feld nach der relativistischen theorie diracs.
\newblock {\em Zeitschrift f{\"u}r Physik}, 69(11):742--764, 1931.

\bibitem{schwinger1951gauge}
Julian Schwinger.
\newblock On gauge invariance and vacuum polarization.
\newblock {\em Physical Review}, 82(5):664, 1951.

\bibitem{dunne2005heisenberg}
Gerald~V Dunne.
\newblock Heisenberg-euler effective lagrangians: basics and extensions.
\newblock In {\em From Fields to Strings: Circumnavigating Theoretical Physics: Ian Kogan Memorial Collection (In 3 Volumes)}, pages 445--522. World Scientific, 2005.

\bibitem{fedotov2023advances}
A~Fedotov, A~Ilderton, F~Karbstein, Ben King, D~Seipt, H~Taya, and Greger Torgrimsson.
\newblock Advances in qed with intense background fields.
\newblock {\em Physics Reports}, 1010:1--138, 2023.

\bibitem{affleck1982monopole}
Ian~K Affleck and Nicholas~S Manton.
\newblock Monopole pair production in a magnetic field.
\newblock {\em Nuclear Physics B}, 194(1):38--64, 1982.

\bibitem{affleck1982pair}
Ian~K Affleck, Orlando Alvarez, and Nicholas~S Manton.
\newblock Pair production at strong coupling in weak external fields.
\newblock {\em Nuclear Physics B}, 197(3):509--519, 1982.

\bibitem{dunne2005worldline}
Gerald~V Dunne and Christian Schubert.
\newblock Worldline instantons and pair production in inhomogenous fields.
\newblock {\em Physical Review D}, 72(10):105004, 2005.

\bibitem{dunne2006worldline}
Gerald~V Dunne, Qing-hai Wang, Holger Gies, and Christian Schubert.
\newblock Worldline instantons and the fluctuation prefactor.
\newblock {\em Physical Review D—Particles, Fields, Gravitation, and Cosmology}, 73(6):065028, 2006.

\bibitem{dumlu2010stokes}
Cesim~K Dumlu and Gerald~V Dunne.
\newblock Stokes phenomenon and schwinger vacuum pair production in time-dependent laser pulses.
\newblock {\em Physical review letters}, 104(25):250402, 2010.

\bibitem{Dumlu_2011}
Cesim~K. Dumlu and Gerald~V. Dunne.
\newblock Complex worldline instantons and quantum interference in vacuum pair production.
\newblock {\em Physical Review D}, 84(12), December 2011.

\bibitem{brezin1970pair}
Edouard Br{\'e}zin and Claude Itzykson.
\newblock Pair production in vacuum by an alternating field.
\newblock {\em Physical Review D}, 2(7):1191, 1970.

\bibitem{popov1972pair}
VS~Popov.
\newblock Pair production in a variable external field (quasiclassical approximation).
\newblock {\em Soviet Journal of Experimental and Theoretical Physics}, 34:709, 1972.

\bibitem{monin2010photon}
A~Monin and MB~Voloshin.
\newblock Photon-stimulated production of electron-positron pairs in an electric field.
\newblock {\em Physical Review D}, 81(2):025001, 2010.

\bibitem{monin2010semiclassical}
A~Monin and MB~Voloshin.
\newblock Semiclassical calculation of photon-stimulated schwinger pair creation.
\newblock {\em Physical Review D}, 81(8):085014, 2010.

\bibitem{torgrimsson2017dynamically}
Greger Torgrimsson, Christian Schneider, Johannes Oertel, and Ralf Sch{\"u}tzhold.
\newblock Dynamically assisted sauter-schwinger effect—non-perturbative versus perturbative aspects.
\newblock {\em Journal of High Energy Physics}, 2017(6):43, 2017.

\bibitem{affleck1979induced}
Ian~K Affleck and Frank De~Luccia.
\newblock Induced vacuum decay.
\newblock {\em Physical Review D}, 20(12):3168, 1979.

\bibitem{voloshin1986particle}
MB~Voloshin and KG~Selivanov.
\newblock On the particle-induced decay of a metastable vacuum.
\newblock {\em Sov. J. Nucl. Phys.(Engl. Transl.);(United States)}, 44(5), 1986.

\bibitem{kuznetsov1997false}
AN~Kuznetsov and PG~Tinyakov.
\newblock False vacuum decay induced by particle collisions.
\newblock {\em Physical Review D}, 56(2):1156, 1997.

\bibitem{demidov2015high}
Sergei Demidov and Dmitry Levkov.
\newblock High-energy limit of collision-induced false vacuum decay.
\newblock {\em Journal of High Energy Physics}, 2015(6):1--33, 2015.

\bibitem{gorsky2006particle}
A~Gorsky and MB~Voloshin.
\newblock Particle decay in false vacuum.
\newblock {\em Physical Review D}, 73(2):025015, 2006.

\bibitem{gorsky2002schwinger}
AS~Gorsky, KA~Saraikin, and KG~Selivanov.
\newblock Schwinger type processes via branes and their gravity duals.
\newblock {\em Nuclear Physics B}, 628(1-2):270--294, 2002.

\bibitem{semenoff2011holographic}
Gordon~W Semenoff and Konstantin Zarembo.
\newblock Holographic schwinger effect.
\newblock {\em Physical review letters}, 107(17):171601, 2011.

\bibitem{sato2013holographic}
Yoshiki Sato and Kentaroh Yoshida.
\newblock Holographic schwinger effect in confining phase.
\newblock {\em Journal of High Energy Physics}, 2013(9):1--12, 2013.

\bibitem{hashimoto2015electromagnetic}
Koji Hashimoto, Takashi Oka, and Akihiko Sonoda.
\newblock Electromagnetic instability in holographic qcd.
\newblock {\em Journal of High Energy Physics}, 2015(6):1, 2015.

\bibitem{sonner2013holographic}
Julian Sonner.
\newblock Holographic schwinger effect and the geometry of entanglement.
\newblock {\em Physical review letters}, 111(21):211603, 2013.

\bibitem{jensen2013holographic}
Kristan Jensen and Andreas Karch.
\newblock Holographic dual of an einstein-podolsky-rosen pair has a wormhole.
\newblock {\em Physical review letters}, 111(21):211602, 2013.

\bibitem{grieninger2023entanglement}
Sebastian Grieninger, Dmitri~E Kharzeev, and Ismail Zahed.
\newblock Entanglement entropy in a time-dependent holographic schwinger pair creation.
\newblock {\em Physical Review D}, 108(12):126014, 2023.

\bibitem{grieninger2023entanglement2}
Sebastian Grieninger, Dmitri~E Kharzeev, and Ismail Zahed.
\newblock Entanglement in a holographic schwinger pair with confinement.
\newblock {\em Physical Review D}, 108(8):086030, 2023.

\bibitem{ghodrati2015schwinger}
Mahdis Ghodrati.
\newblock Schwinger effect and entanglement entropy in confining geometries.
\newblock {\em Physical Review D}, 92(6):065015, 2015.

\bibitem{ritus1988processes}
AI~Nikishov and VI~Ritus.
\newblock Processes induced by a charged particle in an electric field, and the unruh heat-bath concept.
\newblock {\em Zh. Eksp. Teor. Fiz}, 94(3):1--47, 1988.

\bibitem{monin2005monopole}
Alexander~K Monin.
\newblock Monopole decay in the external electric field.
\newblock {\em Journal of High Energy Physics}, 2005(10):109, 2005.

\bibitem{gorsky2024}
Gorsky.A and Pikalov.A.
\newblock Schwinger-like pair production of baryons in electric field.
\newblock {\em Jetp.Letters}, 119:751--757, 2024.

\bibitem{satunin2013width}
Petr Satunin.
\newblock Width of photon decay in a magnetic field: Elementary semiclassical derivation and sensitivity to lorentz violation.
\newblock {\em Physical Review D—Particles, Fields, Gravitation, and Cosmology}, 87(10):105015, 2013.

\bibitem{satunin2015study}
Petr Satunin.
\newblock Study of neutrino decay in a magnetic field within the “worldline instanton” approach.
\newblock {\em JETP letters}, 101(10):657--663, 2015.

\bibitem{robl1952pair}
H~Robl.
\newblock Pair creation in a homogenous magnetic field (in german).
\newblock {\em Acta Phys. Austriaca}, 6:105--118, 1952.

\bibitem{baier2007pair}
VN~Baier and VM~Katkov.
\newblock Pair creation by a photon in a strong magnetic field.
\newblock {\em Physical Review D—Particles, Fields, Gravitation, and Cosmology}, 75(7):073009, 2007.

\bibitem{klepikov1953emission}
NP~Klepikov.
\newblock Emission of photons or electron-positron pairs in magnetic fields.
\newblock {\em Zhur. Eksptl.'i Teoret. Fiz.}, 26, 1953.

\bibitem{tsai1974photon}
Wu-yang Tsai and Thomas Erber.
\newblock Photon pair creation in intense magnetic fields.
\newblock {\em Physical Review D}, 10(2):492, 1974.

\bibitem{ginzburg1965pion}
VL~Ginzburg and GF~Zharkov.
\newblock Pion and beta-ray emission by protons moving in a magnetic field.
\newblock {\em Soviet Physics JETP}, 20(6), 1965.

\bibitem{muller1997decay}
Rainer M{\"u}ller.
\newblock Decay of accelerated particles.
\newblock {\em Physical Review D}, 56(2):953, 1997.

\bibitem{PhysRevLett.87.151301}
Daniel A.~T. Vanzella and George E.~A. Matsas.
\newblock Decay of accelerated protons and the existence of the fulling-davies-unruh effect.
\newblock {\em Phys. Rev. Lett.}, 87:151301, Sep 2001.

\bibitem{vanzella2000weak}
Daniel~AT Vanzella and George~EA Matsas.
\newblock Weak decay of uniformly accelerated protons and related processes.
\newblock {\em Physical Review D}, 63(1):014010, 2000.

\bibitem{matsas1999decay}
George~EA Matsas and Daniel~AT Vanzella.
\newblock Decay of protons and neutrons induced by acceleration.
\newblock {\em Physical Review D}, 59(9):094004, 1999.

\bibitem{wistisen2021transmutation}
Tobias~N Wistisen, Christoph~H Keitel, and Antonino Di~Piazza.
\newblock Transmutation of protons in a strong electromagnetic field.
\newblock {\em New Journal of Physics}, 23(6):065007, 2021.

\bibitem{vanzella2001decay}
Daniel~AT Vanzella and George~EA Matsas.
\newblock Decay of accelerated protons and the existence of the fulling-davies-unruh effect.
\newblock {\em Physical review letters}, 87(15):151301, 2001.

\bibitem{blackburn2020}
T.~G. Blackburn.
\newblock Radiation reaction in electron--beam interactions with high-intensity lasers.
\newblock {\em Reviews of Modern Plasma Physics}, 4:5, 2020.

\bibitem{dolgov2024conversion}
AD~Dolgov and AS~Rudenko.
\newblock Conversion of protons to positrons by a black hole.
\newblock {\em Physics of Particles and Nuclei Letters}, 21(4):865--872, 2024.

\bibitem{gorsky2000tunnelling}
A~Gorsky and K~Selivanov.
\newblock Tunnelling into the randall-sundrum brane world.
\newblock {\em Physics Letters B}, 485(1-3):271--277, 2000.

\bibitem{gorsky2007spontaneous}
AS~Gorsky.
\newblock Spontaneous creation of the brane world and direction of the time arrow.
\newblock {\em Physics Letters B}, 646(4):183--188, 2007.

\end{thebibliography}
\end{document}